\documentclass[12pt]{revtex4}  

\usepackage{amsmath}    
\usepackage{amsfonts}  
\usepackage{amssymb}
\usepackage{dsfont}
\usepackage{mathrsfs}
\usepackage{graphicx}   

\begin{document}

\title{Classical optics representation of the quantum mechanical translation operator via ABCD matrices}

\author{Marco Ornigotti$^1$ and Andrea Aiello$^{1,2}$}
\affiliation{$^1$Max Planck Institute for the Science of Light, G$\ddot{u}$nther-Scharowsky-Strasse 1/Bau24, 91058 Erlangen, Germany} 
\affiliation{$^2$Institute for Optics, Information and Photonics, University of Erlangen-Nuernberg, Staudtstrasse 7/B2, 91058 Erlangen, Germany}
 \email{marco.ornigotti@mpl.mpg.de}  
 
\begin{abstract}
The ABCD matrix formalism describing paraxial propagation of optical beams across linear systems is generalized to arbitrary beam trajectories. As a by-product of this study, a one-to-one correspondence between the extended ABCD matrix formalism presented here and the quantum mechanical translation operator is established.
\end{abstract}


\date{\today}
\maketitle

\section{Introduction}
Rays of light propagate along rectilinear trajectories in air. Therefore, at the generic position $x$ a ray, which may be represented by the linear function $f(x)=a+bx$, is completely determined by a pair of numbers solely: $f(x)$ and $f'(x)$. Such a pair may be represented in a vector-like form as follows:\\
\begin{equation}
\textbf{f}(x)=\left[
\begin{array}{c}
f(x)\\
f'(x)
\end{array}\right].
\end{equation}
The simple linear relation existing between $\textbf{f}(x_1)$ and $\textbf{f}(x_2)$ at two arbitrarily chosen positions $x_1$ and $x_2=x_1+L$, with $L>0$, is usually written in optics textbooks \cite{bornNwolf, hecht, pedrotti, burch} in the following matrix form:\\
\begin{equation}\label{prima}
\left[
\begin{array}{c}
f(x_2)\\ f'(x_2)
\end{array}
\right]= \left[
\begin{array}{cc}
1 & L \\ 0 & 1 \\ 
\end{array}
\right] \left[
\begin{array}{c}
f(x_1)\\  f'(x_1)
\end{array} \right].
\end{equation}
The $2\times 2$ matrix in Eq.\eqref{prima} is known as the ABCD matrix of the optical system (free space, in the present case) and fully characterizes the propagation of rays of light in it.

Literature is rich of examples of ABCD matrices for more complicated optical systems as lenses, planes and curved dielectric interfaces, mirrors, inhomogeneous media with a quadratic index profile et cetera, and combinations thereof \cite{Yariv, siegman}. For complex inhomogeneous media, the trajectory of a ray of light is more complicated than a straight line and cannot be represented anymore by a linear function, except then for a very short distance $x_2-x_1\equiv\delta x_1\ll 1$. In this case, a simple approach like the one given by the ABCD matrices fails to be efficient and one has to embrace a more complicated method capable of dealing with the medium inhomogeneities \cite{refSlice}. However, formally, the ABCD matrix approach can still be used if an appropriate generalization of this method is constructed by observing that the ABCD matrix formalism is nothing but  the consequence of a linearization of the trajectory of a ray of light around the initial point $x_1$ \cite{chaos,chaos2}, namely a Taylor expansion truncated up to and including first order terms. Such a linearization procedure, however, is physically meaningful only for $x_2$ close enough to $x_1$: $x_2=x_1+\delta  x_1$. But what if $x_2$ is no longer close to $x_1$? Does the first order Taylor expansion break down? If so, can such expansion be suitably extended? If higher order terms must be retained, what is their physical meaning?

To answer these question we intend to proceed in a two-step reasoning. Firstly, in Sect. 2, we put on rigorous basis this linearization procedure showing that a $2\times 2$ ABCD matrix is a principal sub-matrix \cite{refMat} of an effective $\infty\times\infty$ matrix describing the full nonlinear dynamics of a curvilinear ray of light.  In Sect. 3 we then  discuss the physical meaning of the proposed generalization scheme, pointing out that the generalized ABCD matrix is nothing but a physical representation of the well known quantum translation operator $e^{L(d/dx)}$ in one dimensional quantum mechanics.

\section{Generalized ABCD matrices for non-rectilinear light propagation}
To begin with, let us first re-derive Eq.\eqref{prima} for an arbitrary linear function $f(x)$ that now we write in the following manner:
\begin{equation}\label{seconda1}
f(x) = a_0 +a_1 x \equiv a_0^i + a_1^i (x-x_i),
\end{equation}
where $x_i$ is an arbitrary point belonging to the domain of the function $f(x)$. If we choose $x_i=0$ then we retrieve the previous expression $f(x)=a+bx$ with $a_0=a$ and $a_1=b$. However, for $x_i\neq 0$, the last equality in Eq.\eqref{seconda1} gives:
\begin{subequations}
\begin{eqnarray}
 a_0^i=a_0+a_1x_i=f(x_i),\\
a_1^i=a_1=f'(x_i),
\end{eqnarray}
\end{subequations}
where, for the sake of simplicity, we have introduced the notation $a_k^i=a_k(x_i)$. Since the point  $x_i$ is arbitrarly chosen, we can pick out a different point $x_j=x_i+L$ and write:
\begin{equation}
f(x)=a_0^i+a_1^i(x-x_i)=a_0^j+a_1^j(x-x_j).
\end{equation}
By equating the factors with the same powers of $x$ at the second and third terms in the equation above, we obtain $a_0^j=a_0^i+(x_j-x_i)a_1^i$ and $a_1^j=a_1^i$. This can relation be rewritten in the following matrix form:
\begin{equation}\label{matrice}
\left [
\begin{array}{c}
a_0^j\\ a_1^j
\end{array}
\right]= \left[
\begin{array}{cc}
1 & x_j-x_i \\ 0 & 1 \\ 
\end{array}
\right] \left[
\begin{array}{c}
a_0^i\\  a_1^i
\end{array} \right].
\end{equation}
This result is equivalent with the one written in Eq. \eqref{prima}, if we identify $a_0^j=f(x_2)$, $a_1^j=f'(x_2)$,  $a_0^i=f(x_1)$ and $a_1^i=f'(x_1)$. Obviously, the formal derivation of Eq.\eqref{prima} via the steps (3-6) it is highly redundant for the linear-function case. However, it has the virtue to be generalizable to the case of non-rectilinear ray propagation.

Now, in order to describe a ray that propagates in an arbitrary inhomogeneous medium we need a generic smooth non-linear function $f(x)$ which can be expanded in a Taylor series around $x=0$ as follows:
\begin{equation}\label{otto}
f(x)=a_0+a_1x+a_2x^2+\cdots .
\end{equation}
For any $x_i\in\mathbb{ R}$ we can write $x=x-x_i+x_i$ and insert this relation into Eq.\eqref{otto} to obtain
\begin{equation}\label{formula}
f(x)=a_0+a_1(x-x_i+x_i)+a_2(x-x_i+x_i)^2+\cdots=\sum_{n=0}^{\infty}a_n^i(x-x_i)^n,
\end{equation}
where the $a_n^i$ coefficient are given by:\\
\begin{subequations}
\begin{eqnarray}
a_0^i&=&f(x_i),\\
a_n^i&=&\frac{1}{n!}\frac{d^nf(x)}{dx^n}\Big|_{x=x_i}=\sum_{k=n}^{\infty}\binom{k}{n}a_k x_i^{k-n}.
\end{eqnarray}
\end{subequations}
Now we can repeat the same reasoning that lead to Eq.\eqref{formula}, but with a different expansion point $x_j=x_i+L$, and write the following equality:
\begin{equation}
\sum_{n=0}^{\infty}a_n^i(x-x_i)^n=\sum_{n=0}^{\infty}a_n^j(x-x_j)^n,
\end{equation}
which simply states the independence of $f(x)$ from the expansion points $x_i$ and $x_j$. By expanding both sides of this equation with the help of the Newton's binomial formula one obtains:
\begin{equation}
\sum_{n=0}^{\infty}  a_n^i  \sum_{k=0}^n\binom{n}{k}x^k(-x_i)^{n-k}=\sum_{n=0}^{\infty}a_n^j\sum_{k=0}^n\binom{n}{k}x^k(-x_j)^{n-k}.
\end{equation}
This expression can be turned into a recursive relation by equating terms with the same power of $x$. Then, for  $k=0$ we have:
\begin{equation}
\sum_{n=0}^{\infty}(-1)^na_n^ix_i^n=\sum_{n=0}^{\infty}(-1)^na_n^jx_j^n,
\end{equation}
which can be rewritten, after isolating the $n=0$ term, as:
\begin{equation}
a_0^j=a_0^i+\sum_{n=1}^{\infty}(-1)^n(a_n^ix_i^n-a_n^jx_j^n).
\end{equation}
For $k=1$ the same operation yields:
\begin{equation}
a_1^j=a_1^i+\sum_{n=2}^{\infty}(-1)^{n-1}(a_n^ix_i^{n-1}-a_n^jx_j^{n-1}).
\end{equation}
This procedure can be iterated for arbitrary values of $k$ thus generating the following recursive relation:
\begin{equation}\label{ricorsione}
a_k^j=a_k^i+\sum_{n=k+1}^{\infty}\binom{n}{k}(-1)^{n-k}(a_n^ix_i^{n-k}-a_n^jx_j^{n-k}),
\end{equation}
with $k=0,1,\cdots,n$. The equation above can be seen as a linear algebraic system relating the variables $a_n^i$ to the quantities $a_n^j$.
This result can be then written in matrix form as follows:
\begin{equation}\label{eq16}
 \left[
\begin{array}{cccc}
b_0^j(0) & b_1^j(0) & b_2^j(0)  & \cdots\\
0 & b_1^j(1) & b_2^j(1)  & \cdots\\
0 & 0 & b_1^j(2) & \cdots\\
\vdots & \vdots & \vdots & \ddots
\end{array}
\right] 
\left[\begin{array}{c}
a_0^j\\
a_1^j\\
a_2^j\\
\vdots\\
\end{array}\right]
=
 \left[
\begin{array}{cccc}
b_0^i(0) & b_1^i(0) & b_2^i(0)  & \cdots\\
0 & b_1^i(1) & b_2^i(1)  & \cdots\\
0 & 0 & b_1^i(2) & \cdots\\
\vdots & \vdots & \vdots & \ddots
\end{array}
\right]
\left[\begin{array}{c}
a_0^i\\
a_1^i\\
a_2^i\\
\vdots\\
\end{array}\right],
\end{equation}
where $\textbf{a}^j=(a_0^j,a_1^j,\cdots)$, $\textbf{a}^i=(a_0^i,a_1^i,\cdots)$ and $b_n^j(k)=\binom{n}{k}(-1)^{n-k}x_j^{n-k}$. If we call $\textbf{B}(j)$ the matrix on the left-side of the previous equation and $\textbf{B}(i)$ the one on the right-side, Eq.  \eqref{eq16} can be written in the compact form:
\begin{equation}
\textbf{B}(j)\textbf{a}^j=\textbf{B}(i)\textbf{a}^i,
\end{equation}
where again the shorthand notations $\textbf{B}(i)=\textbf{B}(x_i)$ and $\textbf{B}(j)=\textbf{B}(x_j)$ are used for the sake of clarity.
Solving for $\textbf{a}^j$ by multiplying on the left both sides of the previous equation by $\textbf{B}(j)^{-1}$ and defining $\textbf{A}=\textbf{B}^{-1}(j)\textbf{B}(i)$, we can write the relation between the vectors $\textbf{a}^j$ and $\textbf{a}^i$ as\\
\begin{equation}\label{matrix_form}
\textbf{a}^j=\textbf{A}\textbf{a}^i.
\end{equation} 
The matrix $\textbf{A}$ is our sought \emph{generalized ABCD} matrix, whose expression is the following:\\
\begin{eqnarray}\label{matrice}
\textbf{A} =\textbf{B}^{-1}(j)\textbf{B}(i)= \left[
\begin{array}{ccccc}
1 & L & L^2 & L^3 & \cdots\\
0 & 1 & 2L & 3L^2 & \cdots\\
0 & 0 & 1 & 3L & \cdots\\
0 & 0 & 0 & 1 & \cdots\\
\vdots & \vdots & \vdots & \vdots & \ddots
\end{array}
\right]\nonumber \\
&\equiv& \textbf{A}(L),
\end{eqnarray}
where $L=x_j - x_i$. Note that this matrix contains the usual (i.e. linear) ABCD matrix defined in Eq.\eqref{prima} as the first $2\times2$ principal sub matrix. This sub matrix verifies the following equality:\\
\begin{equation}\label{abcd2d}
\left[
\begin{array}{cc}
1 & L_1 \\ 0 & 1 \\ 
\end{array}
\right]\left[
\begin{array}{cc}
1 & L_2 \\ 0 & 1 \\ 
\end{array}
\right]=\left[
\begin{array}{cc}
1 & L_1+L_2 \\ 0 & 1 \\ 
\end{array}
\right],
\end{equation}
as can be checked by a straightforward calculation. Simliarly, for the $3\times 3$ principal sub matrix one obtains:
\begin{equation}\label{eq23}
\left[
\begin{array}{ccc}
1 & L_1 &L_1^2 \\ 0 & 1 & 2L_1 \\
0 & 0 & 1\\ 
\end{array}
\right]\left[
\begin{array}{ccc}
1 & L_2 &L_2^2 \\ 0 & 1 & 2L_2 \\
0 & 0 & 1\\ 
\end{array}
\right]
= \left[
\begin{array}{ccc}
1 & L_1+L_2 & (L_1+L_2)^2 \\ 0 & 1 & 2(L_1+L_2) \\
0 & 0 & 1
\end{array}
\right].
\end{equation}
\section{Connection with the translation operator}
The composition properties of the various sub-matrices, as given  for the linear and quadratic order by  Eq. \eqref{abcd2d} and Eq. \eqref{eq23} respectively, have a straightforward physical meaning: they illustrate the fact that propagation across two consecutive distances $L_1$ and $L_2$ can be described as a single propagation along the distance $L_1+L_2$. From a mathematical point of view, this is a signature of the semigroup property of our generalized ABCD matrices \cite{burch}.
With the help of a suitable mathematical software for algebraic manipulation, it is not difficult to verify via explicit $N\times N$ matrix multiplications, that Eq.\eqref{eq23} is valid for arbitrary $N$. Thus, by iteration, one can easily convince oneself that the matrix $\textbf{A}$ satisfies the following relation \cite{hecht}:\\
\begin{equation}\label{eq24}
\prod_{n=1}^N \textbf{A}(L_n)=\textbf{A}\Big(\sum_{n=1}^N L_n\Big).
\end{equation}
The physical implications of this relation are immediately understood: the propagation of the function through the total distance $L_1+L_2+\cdots+L_N$ can be achieved by consecutive propagation across the distances $L_1$, $L_2$, $\cdots$, $L_N$. 

This analogy is not accidental. A closer inspection to Eq.\eqref{matrix_form} reveals in fact that this equation tells us how the value of the function $f(x)$ in a point $x_j$ can be calculated knowing the value of the same function in a point $x_i<x_j$. With this in mind, we can calculate the derivative of $f(x)$ as follows:\\
\begin{eqnarray}
\frac{df(x)}{dx}&=&\lim_{\Delta x\rightarrow 0}\frac{f(x+\Delta x)-f(x)}{\Delta x}=\lim_{x_j\rightarrow x_i}\Big(\frac{\textbf{a}^j-\textbf{a}^i}{x_j-x_i}\Big)\nonumber\\
&=&\lim_{L\rightarrow 0}\Big(\frac{\textbf{A}-\textbf{I}}{L}\Big)\textbf{a}^i \equiv \textbf{D}\textbf{a}^i,
\end{eqnarray}
where we have chosen $\Delta x=x_j-x_i\equiv L$ in order to represent $f(x+\Delta x)$ as $\textbf{a}^j$ and $f(x)$ as $\textbf{a}^i$. Note that this does not cause any loss of generality, since the definition of derivative involves only the concept of neighboring points and, as discussed previously, the quantities $\textbf{a}^{i}$ and $\textbf{a}^j$ represent the value of the function $f(x)$ in two arbitrary neighboring points. Note, moreover, that in the last equality we used Eq.\eqref{matrix_form} to write $\textbf{a}^j$ as a function of $\textbf{a}^i$. Here, $\textbf{D}$ is the matrix representation of the differential operator $d/dx$ \cite{article}\\
\begin{equation}
\textbf{D}=\lim_{L\rightarrow 0}\Big(\frac{\textbf{A}-\textbf{I}}{L}\Big) = \left[
\begin{array}{ccccc}
0 & 1 & 0 & 0 & \cdots\\
0 & 0 & 2 & 0 & \cdots\\
0 & 0 & 0 & 3 & \cdots\\
0 & 0 & 0 & 0 & \cdots\\
\vdots & \vdots & \vdots & \vdots & \ddots
\end{array}\right].
\end{equation}
At this point, it is not difficult to see, via an explicit calculation, that the generalized ABCD matrix is related to the differential operator by the following formula:\\
\begin{equation}\label{questa}
\textbf{A}(L)=\sum_{k=0}^{\infty}\frac{(L\textbf{D})^k}{k!}=e^{L\textbf{D}}.
\end{equation}
Equation \eqref{questa} gives therefore an actual \emph{physical} representation of the well-known translation operator $e^{L(d/dx)}$ \cite{ziman}, such that:
\begin{equation}
e^{L\frac{d}{dx}}f(x)|_{x=0} = \sum_{k=0}^{\infty}\frac{f^k(0)}{k!}L^k=f(L).
\end{equation}
In our case, in fact, the action of the $\textbf{A}$ matrix completely defines the vector $\textbf{F}(x_2)\equiv\textbf{a}^j$ at the point $x_2$ knowing the expression of the $\textbf{F}(x_1)\equiv\textbf{a}^i$ at the point $x_1$, i.e.\\
\begin{equation}
\textbf{F}(x_2)=e^{L\textbf{D}}\textbf{F}(x_1)=\textbf{A}(L)\textbf{F}(x_1).
\end{equation}

\section{Conclusions}
In this work we generalized the concept of ABCD matrix to the case of arbitrary beam trajectories by noticing that the usual ABCD matrix is only the principal $2\times 2$ sub matrix (i.e., the lowest order approximation) of an $\infty\times\infty$ matrix describing the beam dynamics in the general case. A closer inspection of this generalization allowed us to establish a one-to-one connection between the generalized ABCD matrix and the quantum mechanical translation operator in one dimension.

\section*{References}

\end{document}